\documentclass[journal]{IEEEtran}
\usepackage{url}

\pdfoutput=1

\usepackage{cite}

\ifCLASSINFOpdf
   \usepackage[pdftex]{graphicx}
\else
   \usepackage[dvips]{graphicx}
   \DeclareGraphicsExtensions{.eps}
\fi

\usepackage{floatrow}

\begin{document}
\title{Externalization of Packet Processing in Software Defined Networking}

\author{Douglas~Comer, and Adib~Rastegarnia
\thanks{Douglas Comer and Adib Rastegarnia are with the Department
of Computer Science, Purdue University, IN,
West Lafayette, 47906 USA e-mail: comer@cs.purdue.edu, arastega@purdue.edu}% <-this % stops a space
\thanks{}}

\maketitle

\begin{abstract}
Current SDN controllers aggregate all control plane subsystems into a monolithic program. A controller that follows the aggregated approach defines its own set of programming interfaces and services, making application development dependent on a particular SDN controller and restricting portability of management applications across controllers. We propose a new architecture that disaggregates controller functionality and externalizes packet processing, a critical first step towards migrating from a centralized, monolithic design to a decentralized microservice control plane architecture in which SDN controller functions are divided into a smaller, interconnected set.  We argue that dividing a monolithic controller into smaller pieces has advantages.
\end{abstract}

\begin{IEEEkeywords}
Software Defined Networking, Control Plane Disaggregation, Apache Kafka, Message Distribution System, OpenFlow.
\end{IEEEkeywords}

% \fi
\IEEEpeerreviewmaketitle

\section{Introduction}
\IEEEPARstart{T}{he} next generation of network management systems will be based on \emph{Software Defined Networking (SDN)}, an emerging paradigm that breaks the vertical integration of control and data planes to allow software modules that run outside a given network device to configure devices according to a set of network policies. The current SDN paradigm partitions functionality into three broad pieces: a data plane, a control plane, and an application plane \cite{7452335}. The control plane is implemented by an \emph{SDN controller}. Most SDN controllers employ two types of APIs used to communicate with outside entities: a \emph{Northbound (NB)} interface that defines communication between an external management application and control plane software running in the controller, and a \emph{Southbound (SB)} interface that defines communication between the control plane software running in the controller and underlying network devices \cite{8066287}. Early SDN work defined the \emph{OpenFlow} \cite{McKeown:2008:OEI:1355734.1355746} SB API, and the OpenFlow protocol continues to dominate the southbound protocol space. OpenFlow allows a controller to update flow table rules, and to specify associated actions to be performed for each of the flows that pass through a given network device. NB APIs offer a programming abstraction to application developers, allowing them to build network applications, such as load-balancers, traffic engineering systems, firewalls, and network monitoring facilities.  The current software-defined management architecture exhibits several properties that can be considered weaknesses:

\begin{itemize}

\item \textbf{Monolithic and Proprietary}: In the current architectural approach, an SDN controller aggregates all control plane subsystems into a single, monolithic program. Each controller that adopts the aggregated control plane model defines its own services and its own NB interface used to access the services; a programmer must use the controller-specific interface when developing applications and other software for the controller. The approach makes application development dependent on a particular SDN controller and restricts portability of management applications across controllers. Although it enabled vendors to create and market SDN controller products, the monolithic approach does not provide modularity, and does not support non-disruptive updates. In addition, monolithic designs do not provide an easy way to scale to handle larger networks.

\item \textbf{Dependency on the NB API of a specific SDN Controller}: 
The NB APIs offered by SDN controllers such as ONOS\cite{Berde:2014:OTO:2620728.2620744} and OpenDayLight differ. In addition, even in cases where two or more controllers use the same general approach for a NB API (e.g., a RESTful approach), the APIs differ in terms of syntax, naming convention, resource exposed, and so on. The lack of a uniform set of NB APIs means a given management application depends on the NB API of a specific SDN controller, reducing portability. To solve the problem, the Open Networking Foundation (ONF) created a NB Interface Working  Group (NBI-WG) that intended to create a set of standard NB APIs (at multiple levels of abstraction) for all SDN controllers to adopt. Unfortunately, the effort has not produced widely-accepted standardized NB APIs.

\item \textbf{Lack of Reusability among Software Modules}: In the current SDN architectures, the dependency between a management application and a specific type of controller limits reuse of even basic software components in an SDN application. For example, a module that collects topology information, creates/installs flow rules, monitors failures and topology changes, or collects flow rule statistics must be recoded from scratch when porting an application from one SDN controller to another. 

\item \textbf{Lack of External Reactive SDN Applications}: In the current SDN architectures, network programmers use NB APIs to  program network devices \emph{proactively}. That is, an
application installs flow rules to handle all possible cases before traffic arrives.  To exploit the full flexibility of SDN, a system must support a \emph{reactive} approach in which external management applications are informed of changes in the network and react accordingly. Current SDN controllers do not provide mechanisms to inform external applications when conditions change.

\end{itemize}

To overcome the above weaknesses, SDN controllers must be redesigned to allow all services and apps to run outside of the controllers. One of the first steps to achieve that goal consists of outsourcing packet processing (i.e., moving from monolithic services inside a controller to the external processes that can run outside of the controller). Externalization of packet processing helps to achieve the following long-term goals: 
\begin{itemize}
      \item Migrate from a monolithic control plane design to a \emph{microservice control plane} architecture that allows an SDN controller to be divided into a set of many, small interconnected services instead of a single monolithic application. 
      \item Add support for external reactive applications using facilities other than conventional SDN controller APIs. 
    \item Provide a set of core subsystems that applications can use (e.g., topology and flow subsystems) that run outside of the controller and can be accessed from any programming language.
    
    \item Allow programmers to choose an arbitrary programming language when developing SDN applications, providing any requirement to force programmers to use the same language used to implement the controller itself.
\end{itemize}

To achieve the above design goals, we propose to place a message distribution system at the center of the design. We propose using Kafka as an initial choice.
The idea is to execute Kafka subsystem inside the controller, and use the system to forward incoming packets (i.e., packets that arrive from network devices) to external management processes and applications. The rest of this letter is organized as follows; the following section reviews related work. Section \ref{architecture}  presents an  overview  of the proposed architecture  and  explains  its key  components. Section \ref{setup}  explains the  experimental environments  used  in  the paper. Section \ref{results} presents  our  experimental scenarios  and  measurements. Sections \ref{discussion} and \ref{conclusion} discusses open research problems and concludes the paper, respectively.

\section{Related Work}
\textbf{OFtee} \cite{oftee} is a utility tool that sits between and OpenFlow device and an OpenFlow controller to bidirectionally passes, as is, all traffic between the device and the OpenFlow controller.  The main purpose of this utility filter is to allow programmers to develop and execute external SDN applications outside the SDN controller processes without requiring to develop them for a specific SDN controller such as ONOS or OpenDayLight. Specifically, OFtee can be configured to forward copies of OpenFlow packets to third party applications via REST. OFtee module supports an API that can send ``PACKET\_OUT'' messages to a switch port, because most SDN controllers do not support an API that allows applications to emit ``PACKET\_OUT'' packets destined to a specific device. However, OFtee is a proof of concept that cannot be used as a salable solution in production.  \\ 
\textbf{Umbrella} \cite{Comer:2018:UUS:3230718.3233546} is a unified software defined network programming framework  that provides a new set of APIs for implementing of SDN applications, keeping the abstractions independent of the NB APIs used by specific SDN controllers. Umbrella uses a hybrid approach that utilizes both of reactive and proactive approach for managing and programming of SDN networks. To support external reactive based SDN applications, Umbrella uses OFtee as a tool to provide OpenFlow \emph{PACKET\_IN} messages to the external applications.

\section{An Overview of the Proposed Architecture}
\label{architecture}
\begin{figure}
    \centering
    \includegraphics[width=\columnwidth]{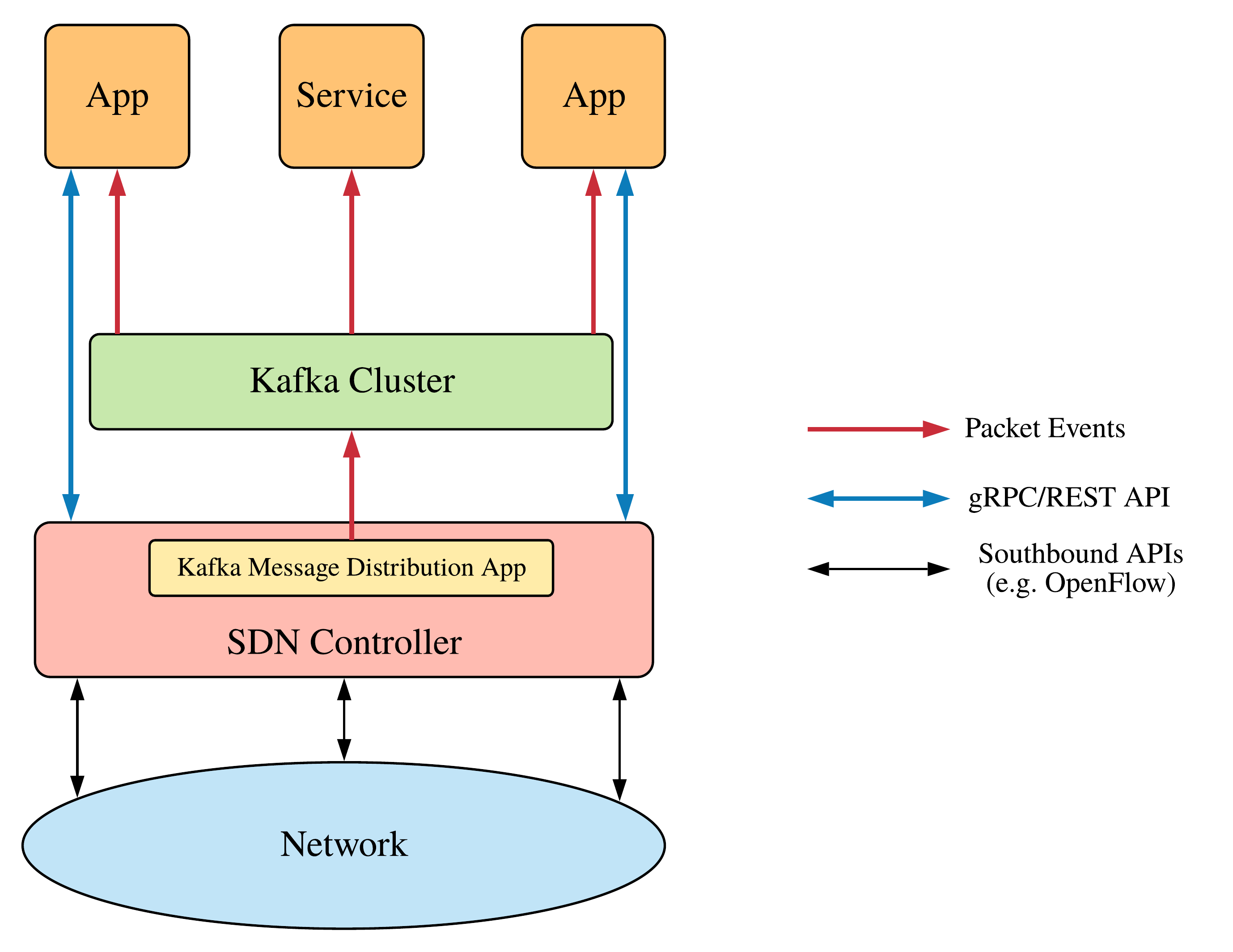}
    \caption{The architecture of the proposed message distribution system}
    \label{fig:1}
\end{figure}

\begin{figure}
    \centering
    \includegraphics[width=\columnwidth]{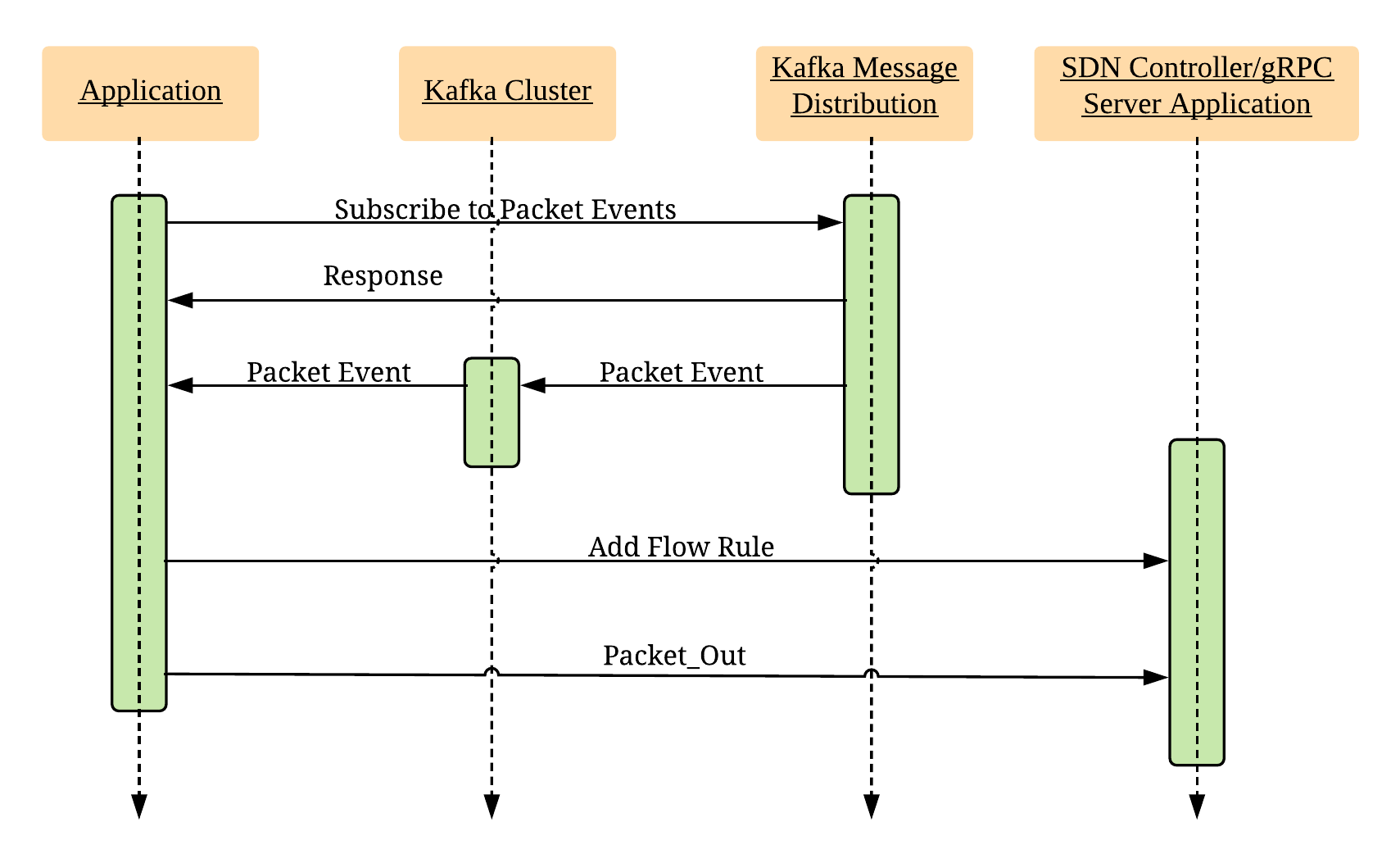}
    \caption{External packet processing and flow rule installation sequence diagram}
    \label{fig:2}
\end{figure}

Figure \ref{fig:1} illustrates the proposed SDN architecture that supports externalized packet processing. The key components are:
\begin{itemize}
    \item \textbf{Kafka Message Distribution App}: The first step in externalizing packet processing consists of adding a mechanism to SDN controllers that provides copies of incoming packets to external management processes. We use a produce-consumer approach to achieve the goal. Our solution is designed around Apache Kafka \cite{kafka} that is an open-source stream-processing software platform developed by the Apache Software Foundation. A Kafka message distribution application listens to the stream of incoming packets (that arrive from network devices), and publish them on a Kafka cluster to be consumed by external management applications and processes. In other words, the message distribution application acts as a producer that pushes data to the brokers on Kafka cluster. 
   
    \item \textbf{Applications and Services}: As Figure \ref{fig:2} illustrates, whenever it needs to receive incoming packets from the devices, a management application subscribes to packet events by sending an HTTP request to the Kafka message distribution app. When it receives a request, the Kafka message distribution system subscribes the requester to the specified topic, and replies with a confirmation. In our implementation, the Kafka message distribution app encodes each packet in a protobuf message, and publishes the result to the Kafka cluster as an array of bytes. Whenever an application receives a packet by consuming it from Kafka cluster, the application must decode and parse the message before processing.  To permit full generality, the system allows an application to return the incoming packet to the pipeline by making a gRPC remote procedure call; the application can use a REST API or gRPC to install flow rules.
    
     \item \textbf{Kafka Cluster}: A Kafka cluster consists of one or more Kafka brokers that run Kafka.
     \end{itemize}
    The following subsection explains two ways the proposed solution can be used to implement at an application and a service outside of an SDN controller.

\subsection{Example uses of the Proposed Solution}
This section uses a reactive management application and a management service to illustrate how the proposed mechanism allows facilities to  be implemented outside of the SDN controller.
    \begin{itemize}
        \item \textbf{Reactive forwarding application}: The application acts in response to packets for which no forwarding rule exists, and installs a new rule for the flow. To implement the application, the application must receive an incoming packet, extract required match fields, generate flow rule(s), install the rules on appropriate network devices, and then return the incoming packet to the network. To install flow rules and return packets to the network, the application can use whatever grpc/REST API northbound interfaces the SDN controller provides. For example, our example implementation uses the controller's REST API to install flow rules, and uses gRPC to return an incoming packet to the network.
        \item \textbf{Topology Discovery Service}: a typical SDN controller provides several built-in services, such as a topology discovery. Some built-in services can be implemented outside of the SDN controller using an arbitrary programming language by using the proposed message distribution system to externalize packet processing. To implement a topology discovery service, for example, one only needs to capture LLDP and ARP packets to derive a mapping of the links between switches and end-hosts.  
    \end{itemize}

\section{Experimental Setup}
\label{setup}
\subsection{Implementation details}
We implemented an early version of the message distribution system using the ONOS \cite{Berde:2014:OTO:2620728.2620744} SDN controller. To further demonstrate our ideas, we customized a version of the Umbrella framework to support a packet event consumer. We also implemented a simple gRPC NB interface in ONOS that can be used by external applications to return incoming packets to the network. In addition, we take advantage of the Umbrella REST API to install flow rules on network switches.

\subsection{SDN Experimental Testbed}
We ran our experiments on our  SDN  testbed  that consists  of 10 OpenFlow  switches that logically define  5 interconnected sites. Each physical switch is divided  into  10  independent  smaller  switches  using virtualized  mode.  Each site follows a Fat-tree network topology as Figure \ref{fig:fat_tree} illustrates. 

\section{Experimental Scenarios and Results}
\label{results}
\begin{figure}
    \centering
    \includegraphics[scale=0.24]{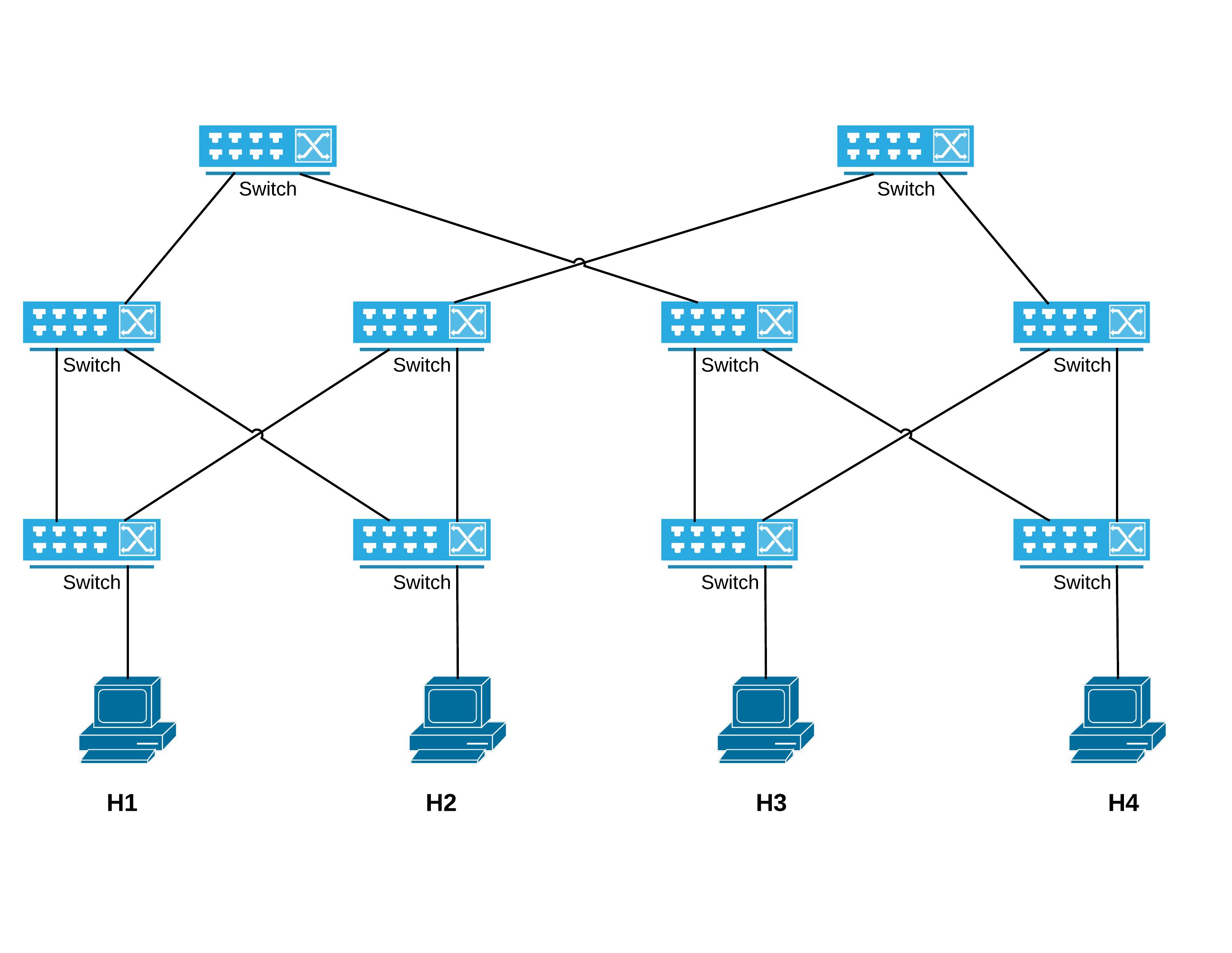}
    \caption{The Fat-Tree network topology in the SDN testbed}
    \label{fig:fat_tree}
\end{figure}

\begin{figure}
    \centering
    \includegraphics[scale=0.65]{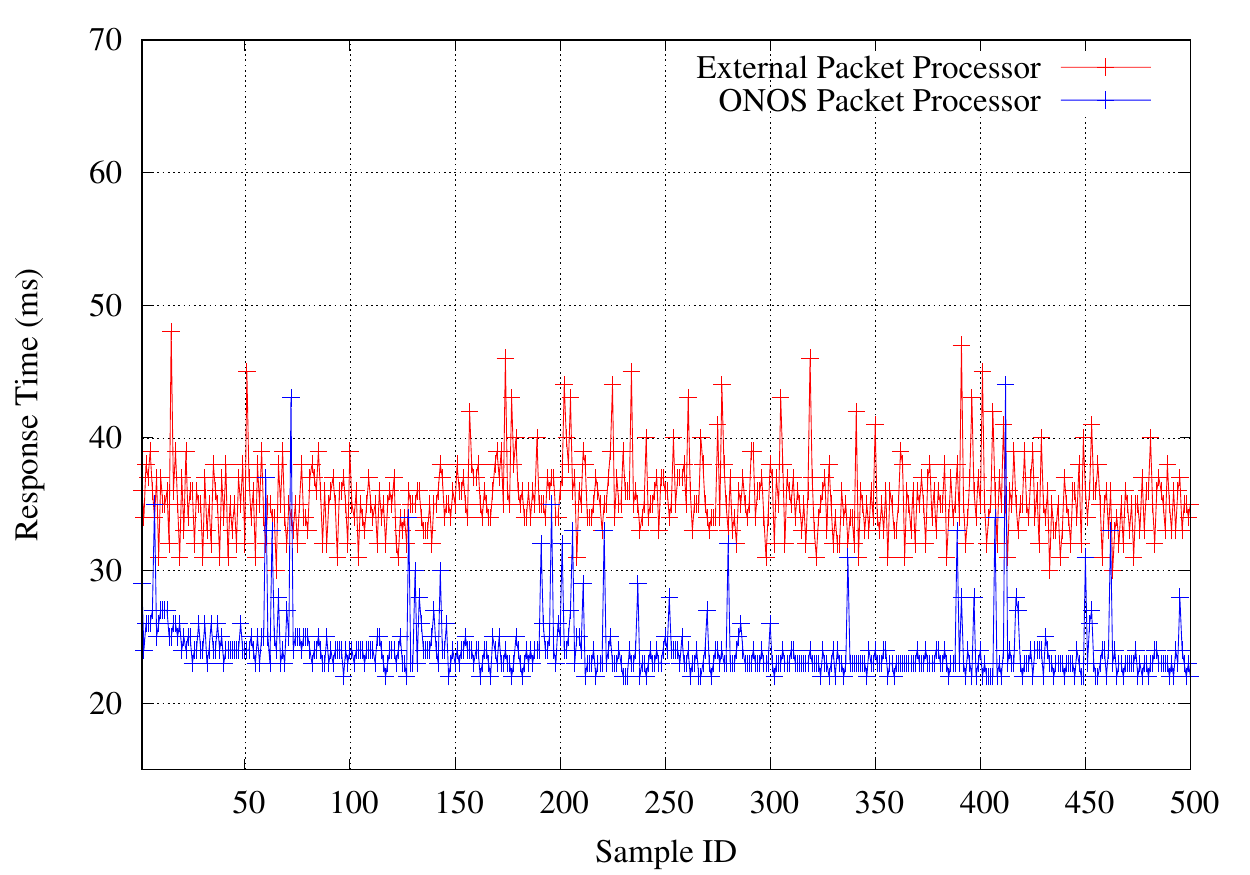}
    \caption{Response time, external packet Processor vs. ONOS packet processor}
    \label{fig:response_time}
\end{figure}

\begin{figure}
    \centering
    \includegraphics[scale=0.65]{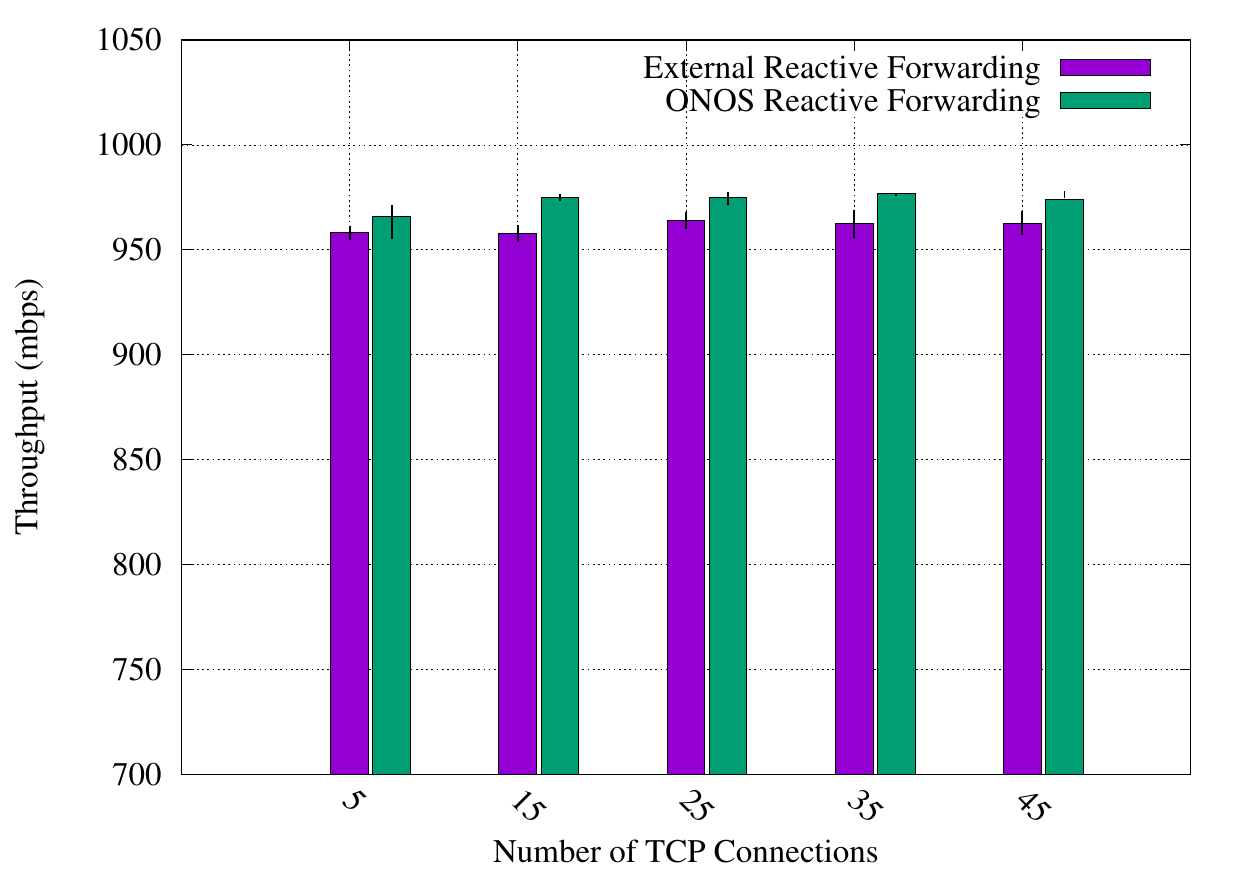}
    \caption{Throughput vs. number of TCP connections for external and ONOS reactive forwarding}
    \label{fig:throughput}
\end{figure}

\begin{itemize}
    \item \textbf{Scenario 1}: The main goal of the first scenario is to compare the amount of time needed to process an incoming packet the packet processor inside ONOS  vs. an external packet processor and to assess the impact on response time. We define the response time as the amount of time that a host needs to send a ping request and receive a reply.  In the standard ONOS packet processing scenario,  whenever a host sends a ping request, each switch along the path sends the incoming packet to the controller, which processes the packet internally and returns it to the pipeline. In the external packet processing scenario, whenever a host sends a ping request, each switch along the path sends the incoming packet to the controller and the Kafka messages distribution application publishes the packet to the Kafka cluster to be processed by an external packet processor.  The external process uses gRPC to return the packet to the pipeline. In this scenario, no flow rule installation is needed to forward packets along the path from one switch to the next switch. We ran the experiment 500 times between hosts H1 and H4 as illustrated in Figure \ref{fig:fat_tree}, and measured the ping response time.  Figure \ref{fig:response_time} summarizes the results, which show the response time ranges from 24ms to 35ms on average, which is insignificant when compared with benefits of externalization of packet processing.  In addition, the delay can be potentially reduced in production if we tune Kafka configuration parameters and gRPC according to our environmental setup; such optimization is outside the scope of this letter. 
    
    \item \textbf{Scenario 2}: In the second scenario, we compare an external reactive forwarding application with an ONOS reactive forwarding application to assess the impact that externalization and the use of a REST API for flow rule installation has on throughput. In both experiments, we use a time-out of 10 seconds to remove flow rules (i.e., flow rules disappear from switches every 10 seconds). We use the iperf3 tool to initiate various numbers of TCP connections that carry emulated web traffic for 150 seconds between two end-hosts. As the results in Figure \ref{fig:throughput} show, the ONOS reactive forwarding application slightly performs better than the external reactive forwarding, mostly because the ONOS REST API that external applications use to install flow rules is slow.
\end{itemize}

\section{Discussion and Future Work}
\label{discussion}
The proposed architecture introduces new research questions that need to be investigated: 
\begin{itemize}
    \item External apps use NB interfaces such as the REST API that current SDN controllers provide to the programmers for installing, removing, and updating flow rules. A REST API  that is not an efficient approach for installing of flow rules in external reactive-based applications because it introduces a notable delay. Replacing the REST APIs with a new NB interface, such as a gRPC-based NB interface, can be considered as a potential solution to speed up flow rule installation in external reactive-based applications. The proposed architecture opens new areas of research in designing of new NB interfaces for SDN controllers to meet new requirements.  
    \item Another research question that need to be studied is where the filtering of packets should be done. We have three options: 
    \begin{itemize}
        \item \textbf{Client side filtering:} In the client side filtering, all of the packets will be sent to the external apps and application are responsible to filter packets according to their requirements.  
        \item \textbf{Server side filtering}: In this option, the Kafka cluster must be equipped with a filtering mechanism to filter packets before providing them to the consumers. 
        \item \textbf{Controller side filtering}: In this option, a client subscribes to specific types of packet and the controller filters packets based on type and publish them on Kafka cluster. 
    \end{itemize}
    Each of the above solutions has its own advantages and disadvantages that need to be studied comprehensively. 
\end{itemize}

\section{Conclusion}
\label{conclusion}
In this letter, we present an architecture using Apache Kafka and gRPC to externalize packet processing that is the first step towards disaggregating the SDN control plane. Externalization of packet processing gives us the flexibility to migrate from a monolithic control plane design to a microservice  control plane architect and split an SDN controller to a set of smaller and interconnected services. We showed that externalization of packet processing introduces some overhead that is negligible in some cases or potentially can be reduced using better technologies or optimization techniques.

\ifCLASSOPTIONcaptionsoff
  \newpage
\fi

\bibliographystyle{IEEEtran}
\bibliography{ref.bib}

\end{document}